\documentclass[twocolumn,preprintnumbers,amsmath,amssymb]{revtex4}
\usepackage{xspace}
\usepackage{graphicx}% Include figure files
\usepackage{dcolumn}% Align table columns on decimal point
\usepackage{bm}% bold math
\usepackage{hyperref}% bold math

\def\beq{\begin{equation}}
\def\eeq{\end{equation}}

\def\beqa{\begin{eqnarray}}
\def\eeqa{\end{eqnarray}}

\newcommand{\roughly}[1]%
    {{\mathrel{\raise.3ex\hbox{$#1$\kern-.75em\lower1ex\hbox{$\sim$}}}}}

\newcommand{\bea}{\begin{eqnarray}\begin{aligned}}
\newcommand{\eea}{\end{aligned}\end{eqnarray}}

\newcommand{\lp}{\left(}
\newcommand{\rp}{\right)}
\newcommand{\order}[1]{\mathcal{O}\lp#1\rp}
\newcommand{\abs}[1]{\left| #1\right|}

\begin{document}

\title{Detecting Hidden Particles with MATHUSLA}

\author{Jared A.~Evans}
\email{jaredaevans@gmail.com}
\affiliation{Department of Physics, University of Illinois at Urbana-Champaign, Urbana, IL 61801, USA}
\affiliation{Department of Physics, University of Cincinnati, Cincinnati, Ohio 45221, USA}

%\pacs{12.60.Fr,14.40.Df,14.40.Nd,14.80.Ec}

\begin{abstract}
A hidden sector containing light long-lived particles provides a well-motivated place to find new physics.  The recently proposed MATHUSLA experiment has the potential to be extremely sensitive to light particles originating from rare meson decays in the very long lifetime region.  In this work, we illustrate this strength with the specific example of a light scalar mixed with the standard model-like Higgs boson, a model where MATHUSLA can further probe unexplored parameter space from exotic Higgs decays.  Design augmentations should be considered in order to maximize the ability of MATHUSLA to discover very light hidden sector particles.
 \end{abstract}

\maketitle

% =============================================================================
\section{\label{sec:intro}Introduction}
% =============================================================================

 The Large Hadron Collider (LHC) continues to collect data and place impressive constraints on new physics from an immense array of possible models.  Despite the plethora of LHC searches for new physics, there have been no new elementary particles discovered since the standard model-like Higgs boson \cite{Aad:2012tfa,Chatrchyan:2012ufa}.    Most searches for new physics have focused on prompt signatures, but the search program for long-lived particles,  those which are produced then propagate some macroscopic distance before decaying, is known to have significant gaps, see  e.g.~\cite{Evans:2016zau,Coccaro:2016lnz}.  Shoring up the gaps in the long-lived particle program is a goal with immediate importance in order to ensure no discovery is missed at the LHC.
 
 One of the most motivated sources for long-lived particles are hidden sectors that only very weakly couple to the standard model via either high dimension operators or very small couplings (for a review, see \cite{Alexander:2016aln}).  A long lifetime for the lightest hidden sector particle due to this feeble connection allows for the new physics to have evaded detection at the LHC and many precision experiments.  Hidden sector models are well-motivated and have been used to explain a wide variety of outstanding deficiencies in the standard model, such as naturalness \cite{Chacko:2005pe,Burdman:2006tz}, dark matter \cite{Tulin:2013teo,Martin:2014sxa,Krnjaic:2015mbs},  inflation \cite{Bezrukov:2009yw}, $m_\nu$ \cite{Dev:2016vle}, the proton radius puzzle \cite{TuckerSmith:2010ra,Barger:2010aj,Batell:2011qq}, and the $(g-2)_\mu$ anomaly \cite{Zhou:2001ew,Pospelov:2008zw}.  The light particles of these hidden sectors can potentially be produced in a variety of ways, including rare meson decays and exotic Higgs decays \cite{Curtin:2013fra}.  One of the most minimal hidden sectors contains a new scalar, $S$, coupled to the standard model via, $\epsilon \left|S\right|^2 H^\dagger H$.  This scalar mixes slightly with the standard model-like Higgs, and provides a simple target for new physics searches  \cite{Bezrukov:2009yw,Clarke:2013aya,Krnjaic:2015mbs,Flacke:2016szy}. 
 
 The proposed MATHUSLA experiment \cite{Chou:2016lxi} (MAssive Timing Hodoscope for Ultra Stable neutraL pArticles) has the potential to access extremely long-lived particles by living symbiotically off of the collisions from the existing LHC program.  MATHUSLA would be an enormous, mostly empty box ($\sim200\times200\times20$ m), containing instrumentation for tracking and vetoing, on the surface roughly 100 m above one of the LHC general purpose detectors (here, assumed to be ATLAS).   Ultra-long-lived particles produced in collisions at ATLAS could traverse the $\order{100 \mbox{ m}}$ of rock without interacting and then decay within this new detector.  MATHUSLA could have background-free detection for long-lived particles with mass $\gtrsim1$ GeV and possibly below \cite{Chou:2016lxi}.   As an enormous number of mesons are produced at LHC, light particles produced in these decays have the potential to be seen by MATHUSLA.   Other experiments have been proposed to access hidden sectors, include the SHiP (Search for Hidden Particles) beam dump \cite{Anelli:2015pba},  the far-forward  FASER \cite{Feng:2017uoz}, and CODEX-b \cite{Gligorov:2017nwh} in the LHCb hall.
 
In this letter, we illustrate that MATHUSLA could be unprecedentedly sensitive to particles with extremely long lifetimes produced in rare meson decays.  In section \ref{sec:meson}, we show that for particles with long lifetimes MATHUSLA exceeds SHiP in sensitivity to $B$ decays, and can be competitive with SHiP for kaon decays if somewhat low energy, $E\sim 200$ MeV, states can be observed, and the backgrounds can be sufficiently controlled.   As a particular case study, we illustrate MATHUSLA's sensitivity to light, Higgs-mixed scalars, first detailing the model in section \ref{sec:hmscalars}, before comparing the potential sensitivity of MATHUSLA with that of other experiments in section \ref{sec:MATHUSLA}.  Conclusions are presented in section \ref{sec:conclusion}.

 % =============================================================================
\section{Hidden Particles in Meson Decays}
\label{sec:meson}
% =============================================================================

To illustrate MATHUSLA's strength at long lifetimes, we compare it to the proposed SHiP beam dump experiment \cite{Anelli:2015pba} within the long lifetime regime in this section.

The rare decays of mesons produced in 14 TeV LHC collisions could yield light, hidden sector particles that would travel a few hundred meters before decaying within the MATHUSLA detector volume.   The LHC production rate for mesons, especially low energy mesons, is fairly uncertain.  To get a quantitative measure of this, we generate $b\bar b$ production in Pythia 8.223 \cite{Sjostrand:2014zea}, and finely bin the outgoing $B$-meson states in energy and angular distributions.  This data is weighted by the total LHC $b\bar b$ cross-section as determined in Pythia, $\sigma_{b\bar b} \approx 0.38$ mb, which is very conservative relative to the 13 TeV LHCb measurement of 0.6 mb \cite{Aaij:2016avz}.  A similar procedure is used to estimate the distributions and rate for kaons, but the kaons, with $c\tau\sim 10$ m, are additionally weighted by the requirement that they decay before reaching the ATLAS calorimeter.  The kaon cross-section is determined by generating soft QCD processes in Pythia with a total cross-section of 0.1 barns (in excellent agreement with TOTEM and ALFA \cite{Antchev:2013paa,Aaboud:2016ijx}).  

For ease of presentation, the mesons are decayed as $B\to X K$, $K_L\to X \pi^0$, and $K^\pm\to X \pi^\pm$, where $X$ is a hidden sector particle.  Other two-body decays, e.g., $K^+\to\mu^+X$ would typically not affect the kinematics greatly.  For different hidden particle lifetimes, the number of decays within MATHUSLA can be computed.   The distribution of particles delivered to SHiP are determined with Pythia 8.223 for a 400 GeV proton beam launched into a fixed-target yielding $N_{b}=6.2\times10^{13}$ and $N_{K}=6\times10^{19}$ \cite{Anelli:2015pba}.  SHiP has a 50 m decay volume beginning 64 m from the target \cite{Anelli:2015pba}, with an elliptical detector of 5.0 (2.5) major (minor) axis.  We model the iron hadronic absorber by requiring kaons to decay before propagating one nuclear interaction length.

\begin{figure}[!t]
\begin{center}
\includegraphics[scale=0.69]{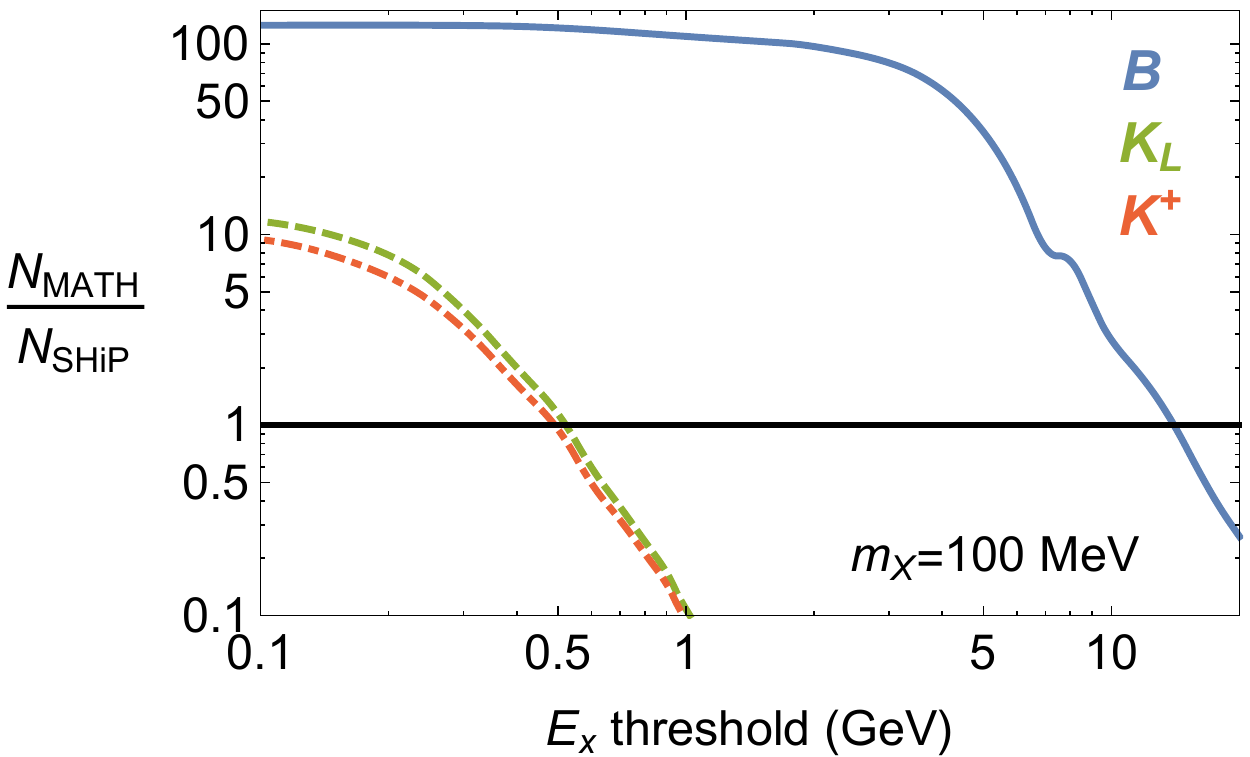}
\end{center}
\caption{The ratio of the total number of long-lifetime hidden sector particles $X$ ($m_X=100$ MeV) above a minimum energy threshold $E_X$ delivered to MATHUSLA in 3 ab$^{-1}$ of LHC data over the total delivered to SHiP (of any energy) for $2\times 10^{20}$ protons on target for the two-body decays of $B$-mesons (blue, solid), $K_L$ (green, dashed), and $K^\pm$ (red, dot-dashed).}  
\label{fig:MvS}
\end{figure}

For large lifetimes, $c\tau\gtrsim 1$ km, the ratio of the number of particles delivered to the two experiments approaches a constant value.  In Fig.~\ref{fig:MvS}, we show the ratio of accepted particles in the two experiments as a function of the minimum energy needed for detection in MATHUSLA.  At these large lifetimes, this is solely a geometric argument independent of the meson decay rate into $X$.   We present result for $m_X=100$ MeV, but for masses that are a significant fraction of the parent meson, the values are typically a little larger (although, flat for energies below $m_X$), with the notable exception of masses near $m_B$, where the smaller number of hidden particles reaching MATHUSLA from the decays of far forward, low energy $B$-mesons reduces sensitivity.  While MATHUSLA has the potential to exceed SHiP at large lifetimes, SHiP is more sensitive to states with $\gamma c\tau\sim50$ m due to the exponential decay rate and larger distance to MATHUSLA.  Although we model two-body decays, similar conclusions may be drawn for three- or more-body decays.  

% =============================================================================
\section{Higgs-Mixed Scalars}
\label{sec:hmscalars}
% =============================================================================

\begin{figure}[t]
\begin{center}
\includegraphics[scale=0.67]{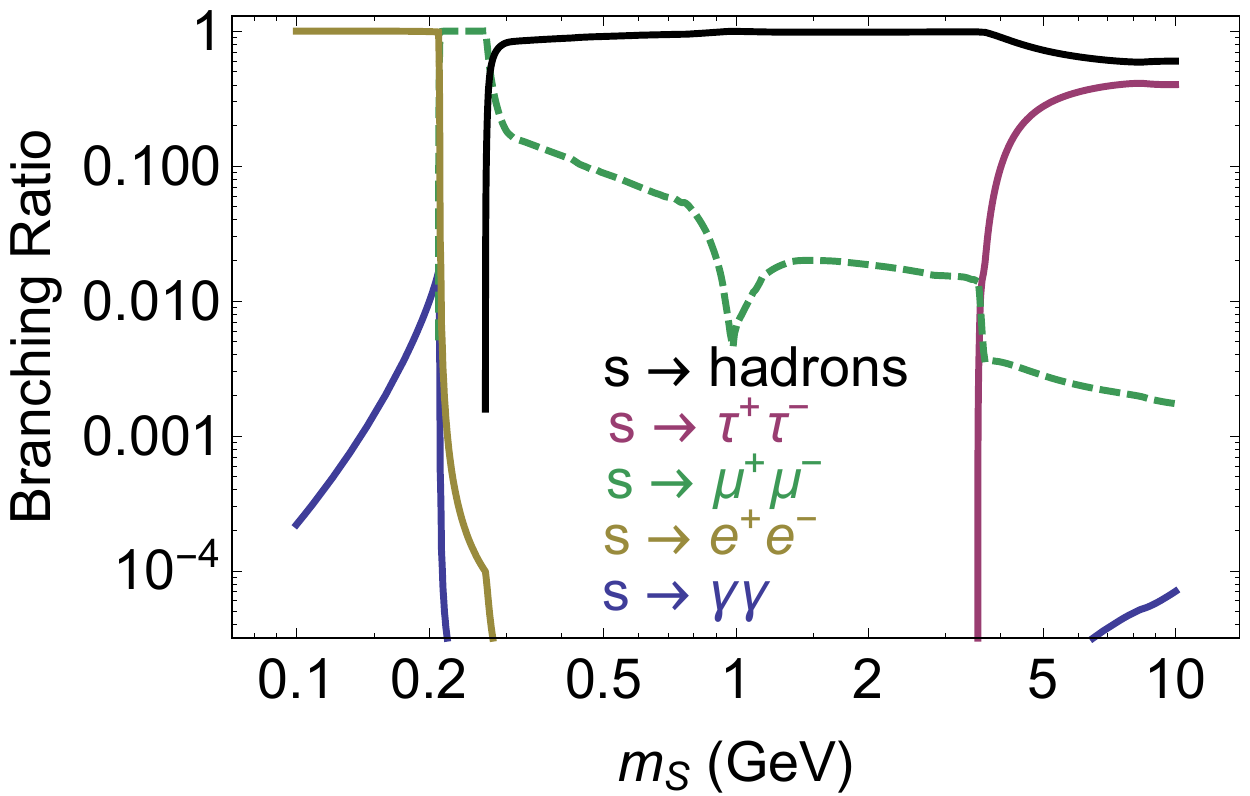}
\end{center}
\caption{Scalar branching ratios in the light hadron region used in this work. For masses below $\sim 1.4$ GeV, we use the results of \cite{Donoghue:1990xh}. We use a smooth extrapolation up to $2m_D$.  The muon branching ratio is shown with the green, dashed curve.}  
\label{fig:ScalarBR}
\end{figure}

To illustrate the potential sensitivity of MATHUSLA with a specific example, we consider light, Higgs-mixed scalars.  A useful simple model to parameterize a Higgs-mixed scalar is
\beq\label{eq:SH}
\mathcal L = \mathcal L_{kin} + \frac{\mu_s ^ 2}{2} S^2-\frac {\lambda_s} {4!} S ^ 4 + \mu ^ 2 |H |^ 2-\lambda |H | ^ 4-\frac{\epsilon}{2}S ^ 2 |H | ^ 2,
\eeq
where $H$ is mostly aligned with the observed, standard model-like Higgs field, and $S$ is a new, real scalar field.  Both the scalar and the Higgs acquire vacuum expectation values, 
\beq
v_s^2 =  6 \frac{\mu_s^2}{\lambda_s }-2 \frac{\epsilon\mu^2}{\lambda \lambda_s }+\order{\epsilon^2},  \hspace{3mm}  v_h^2 = \frac{\mu^2}{\lambda }-3  \frac{\epsilon\mu_s^2}{\lambda \lambda_s }+\order{\epsilon^2}, 
\eeq
resulting in physical states with masses of
\beq
m_s^2 =  \frac 13\lambda_s v_s^2+\order{\epsilon^2},  \hspace{3mm}  m_h^2 =  2\lambda v_h^2+\order{\epsilon^2}. 
\eeq
These states are slightly mixed, with most relevant phenomenological quantities dictated by the mixing angle,
\beq
\tan\theta \approx \frac{\epsilon v_h v_s}{m_h^2-m_s^2} +\order{\epsilon^3}.
\eeq
The standard model-like Higgs state is assigned the values of $m_h=125$ GeV, $v_h=246$ GeV, and total width set to $\Gamma_{h,SM} =4.15$ MeV.  The light scalar's coupling to standard model states is simply
 \beq
 \sin\theta \frac{m_f}{v_h} s f\bar f.
 \eeq
 Assuming that there are no states within the hidden sector lighter than half the scalar mass, the branching ratios of the scalar into standard model particles are the same as those of a standard model-like Higgs boson of the same mass, while the width is simply $\Gamma_s=\sin^2\theta \Gamma_h(m_s)$.  Unfortunately, there is an enormous degree of uncertainty regarding the branching ratios of light scalars with masses in the 0.5--4 GeV region (see \cite{Clarke:2013aya} for an in-depth discussion).   In this region, we follow \cite{Donoghue:1990xh} up to 1.4 GeV and use a smooth interpolation up to the charm threshold.  This implementation is shown in  Fig.~\ref{fig:ScalarBR}.

% =============================================================================
\subsection{Light Scalars in Meson Decays}
\label{sec:Mesons}
% =============================================================================

Higgs-mixed scalars can be emitted in decays of B-mesons (D-mesons and kaons have much smaller branching ratios into a Higgs-mixed scalar and will be neglected here).  Top-loop contributions dominate the partial width, which yield a branching ratio of \cite{Grinstein:1988yu, Gligorov:2017nwh}
\bea
\frac{\mbox{BR}(B\to sX_s)}{\mbox{BR}(B\to X_c e \nu_e)} &= \frac{27\sqrt 2 G_Fm_t^4}{64\pi^2 {\Phi}m_b^2} \!  \abs{\!\frac{V_{ts}^* V_{tb}}{V_{cs}}\!}^{2}\!\!\! \lp\!1\!-\!\frac{m_s^2}{m_b^2}\!\rp^{\!\!2}\sin^2\!\theta \\
\Rightarrow \mbox{BR}(B&\to sX_s) \approx 6.2  \left(1-\frac{m_s^2}{m_B^2}\right)^2  \sin^2\theta.
\label{eq:firstBR}
\eea
where $\Phi\approx 0.5$ \cite{Lenz:2014jha} is a phase space factor for the semi-leptonic decay.  This inclusive branching fraction is inaccurate near $m_s\sim m_B-m_K$ due to the small number of kinematically available exclusive final states.

% =============================================================================
\subsection{Light Scalars in Higgs Decays}
\label{sec:Higgs}
% =============================================================================

The Lagrangian in Eq.~\ref{eq:SH} will also induce an $h\to ss$ decay \cite{Curtin:2013fra,Chang:2016lfq}, however, this decay depends on an additional free parameter from Eq.~\ref{eq:SH} beyond $\sin\theta$ or $m_s$, expressed below by $\lambda_s$.  The partial width for $h\to ss$ is
 \beq
 \Gamma(h\to ss) = \frac{\lambda_s \sin^2\theta m_h^3}{48\pi m_s^2} \lp1 +2\frac{m_s^2}{m_h^2} \rp^2\sqrt{1 - 4\frac{m_s^2}{m_h^2}}.
 \eeq
As $\lambda_s$ gets very small (equivalently, as $v_s$ gets much larger than $m_s$), this branching ratio can grow arbitrarily small.  However, demanding a perturbative $\lambda_s$ ($<16\pi^2$) enforces a maximum branching ratio for $h\to s s$, as a function of $\sin\theta$ and $m_s$, of
\beq
\mbox{BR}(h\to ss) < \frac{\pi \sin^2\theta m_h^3}{3 m_s^2 \Gamma_{h,tot}} \! \lp1 +2\frac{m_s^2}{m_h^2} \rp^2\!\!\sqrt{1 - 4\frac{m_s^2}{m_h^2}}.
\label{eq:BRmax}
\eeq

% =============================================================================
\section{Light Scalars at MATHUSLA}
\label{sec:MATHUSLA}
% =============================================================================

Light scalars produced in the rare decays of $B$-mesons at the 14 TeV LHC could decay within the MATHUSLA detector volume \cite{Fradette:2017sdd,Dev:2017dui}.   We follow the procedure used in section \ref{sec:meson} to derive constraints on scalars produced in the decays of $B$-mesons, but consider the full range of hidden scalar masses and mixing angles.  We require that the scalars decaying within MATHUSLA have $E_s>2$ GeV, which is consistent with the energy thresholds proposed in \cite{Curtin:2017izq}.  Contributions from energetic kaons that decay before reaching the calorimeter contribute only a small correction to the number of scalars delivered, and require more aggressive assumptions about the low energy thresholds of MATHUSLA. 

\begin{figure}[!t]
\begin{center}
\includegraphics[scale=0.69]{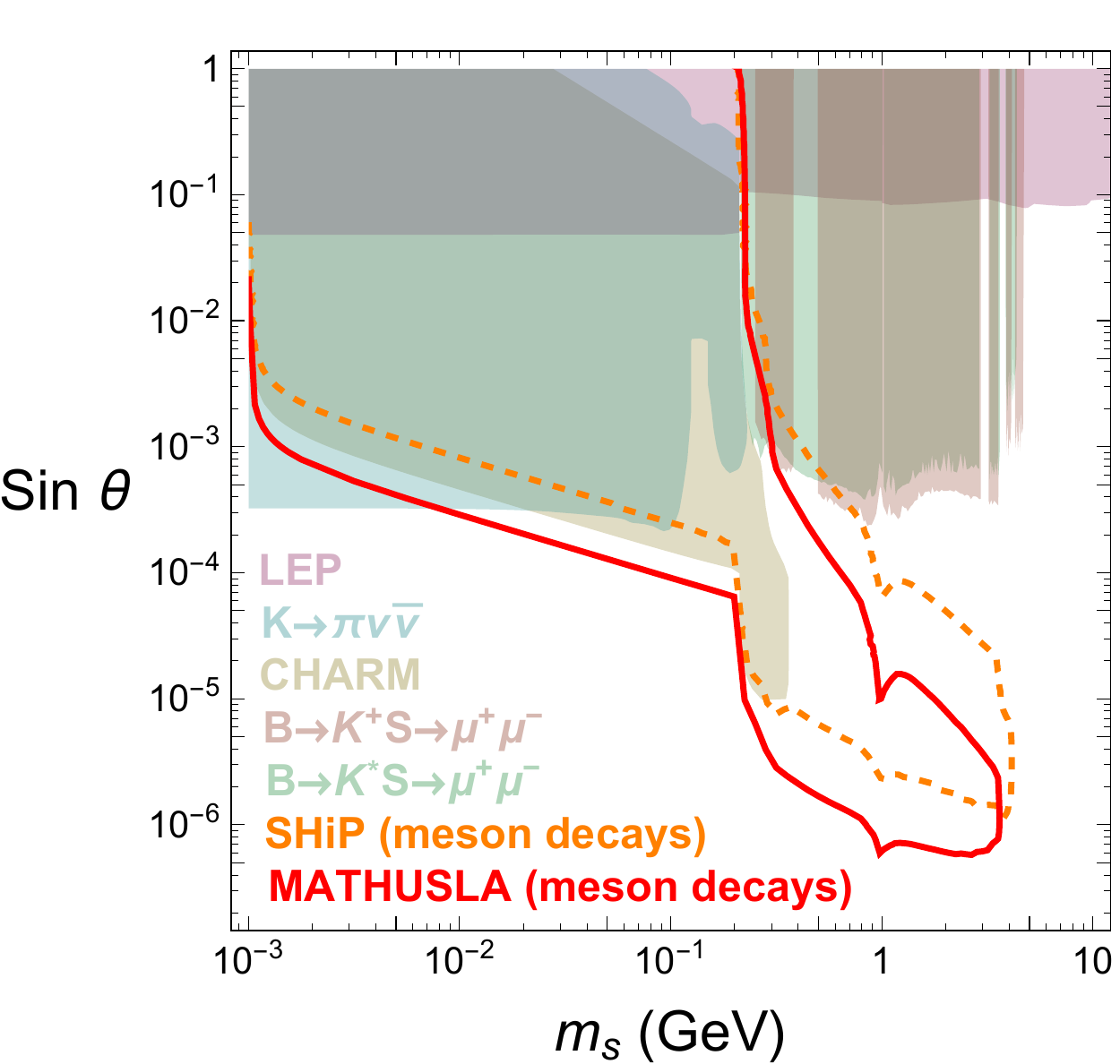}
\end{center}
\caption{Current and projected constraints on light Higgs-mixed scalars.   The projected limits from $B$-meson decays at MATHUSLA are shown by the solid red curve assuming the experiment is background-free.  The projected limits from the SHiP experiment \cite{Anelli:2015pba} are shown by the orange, dashed line.  Other current limits \cite{Acciarri:1996um,Buskulic:1993gi,Bergsma:1985qz,Aaij:2015tna,Aaij:2016qsm,Artamonov:2009sz} are described in the text.}  
\label{fig:MATHNoExo}
\end{figure}

 In  Fig.~\ref{fig:MATHNoExo}, we illustrate by the solid red contour the region of parameter space where four scalars with $E_s>2$ GeV decay within the MATHUSLA volume.  If the experiment is relatively background-free and possesses a signal efficiency of $0.75$, this would correspond to roughly the 95\% confidence level exclusion in the absence of new physics.   Alongside this potential reach, we show current constraints on the parameter space from LEP Higgs searched (light red) \cite{Acciarri:1996um,Buskulic:1993gi}, the CHARM beam dump (gold) \cite{Bergsma:1985qz}, rare $B$ decays at LHCb (light green and brown) \cite{Aaij:2015tna,Aaij:2016qsm}, and $K^\pm \to \pi^\pm +\mbox{invisible}$ at E949 \& E787 (light blue) \cite{Artamonov:2009sz}.    We also show the projected sensitivity at the proposed SHiP experiment \cite{Anelli:2015pba,Alekhin:2015byh} in dashed orange.  In all cases, quoted limits were recalculated using the branching ratios and widths presented in Eq.~\ref{eq:firstBR} and in  Fig.~\ref{fig:ScalarBR}.  
 
 \begin{figure}[!t]
\begin{center}
\includegraphics[scale=0.69]{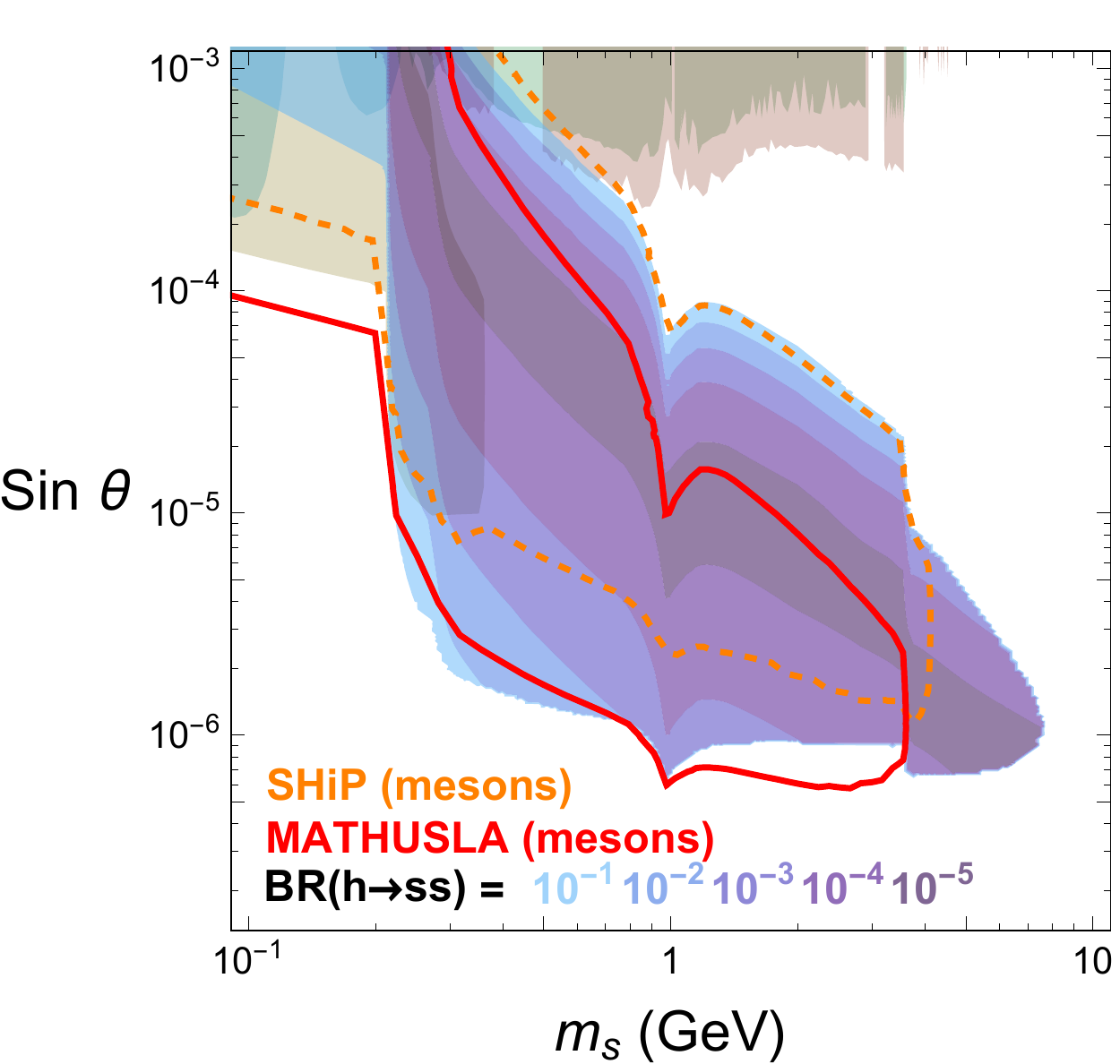}
\end{center}
\caption{Purple contours show the projected sensitivity of MATHUSLA to exotic Higgs decays.  Contours are shown for branching ratios of 10\% to 0.001\% in steps of factors of 10.   Model restrictions on the maximum allowed branching ratio (Eq.~\ref{eq:BRmax}) influence the shape of the contours at higher masses.  The meson decay and other constraints are as in figure \ref{fig:MATHNoExo}.}  
\label{fig:MATHExo}
\end{figure}
 
MATHUSLA also has the potential to detect new light particles produced in exotic Higgs decays \cite{Curtin:2013fra}.  In order to model these exotic Higgs decays, we simulate Higgs production and decay into two scalars with Pythia 8.  As before, these are binned in $m_s$, $E_s$, and angular distributions.  The cross-section is fixed to the total 14 TeV Higgs production cross-section of 62.6 pb \cite{deFlorian:2016spz,Anastasiou:2016cez}.  The absence of scalars delivered to MATHUSLA can bound the $h\to ss$ branching ratio, limits achievable with 3 ab$^{-1}$ are shown with blue-purple contours in  Fig.~\ref{fig:MATHExo}.   At higher masses, the maximum allowed branching ratio (Eq.~\ref{eq:BRmax}) falls below the observable level, sculpting the shape of the sensitivity.  By searching for boosted scalars from exotic Higgs decays, MATHUSLA can probe parameter space beyond the $B$-meson decay constraints.\\

% =============================================================================
\section{Discussion and Conclusions}
\label{sec:conclusion}
% =============================================================================

The MATHUSLA experiment provides a tremendous opportunity to probe light, hidden sectors in the ultra-long-lifetime regime.  Unlike a fixed target experiment, MATHUSLA would also provide sensitivity to scalars produced in exotic Higgs decays, accessing additional regions of parameter space that have no other current experimental prospects.  The ability of MATHUSLA to constrain light particles originating from kaon decays is conditional on it reliably accepting soft signal events and discriminating these from backgrounds.   

Due to the extremely long lifetimes, very light, weakly coupled scalars ($m_s<2m_\mu$ and $\sin\theta\lesssim3\times10^{-4}$) are extremely difficult to discover.  NA62 \cite{NA62TDR,Fantechi:2014hqa} will likely be able to access some of this parameter space, as some theory studies have estimated \cite{Krnjaic:2015mbs,Flacke:2016szy}.  However, the decay in-flight design of NA62 (as opposed to the stopped kaon design of E949, E787, and ORKA \cite{Worcester:2013aje}) reduces the ability of the experiment to distinguish hard pions in the $K^+\to \pi^+s$ decay from the rare muon misidentified as a pion in $K^+\to\mu^+\nu$ decays \cite{Worcester:2013aje}, which could result in a substantially larger background in the light scalar region \cite{Fantechi:2014hqa}.    MATHUSLA is uniquely capable of probing this long lifetime region that lacks any other current experimental prospects.  

 Although light scalars are a well-motivated and interesting example where hidden sector particles are produced in rare meson decays, there are other interesting, long-lived, beyond the standard model particles that could be kinematically accessible in rare $B$-meson or kaon decays, such as light vectors \cite{Pospelov:2008zw}, light right-handed neutrinos \cite{Abazajian:2012ys}, or light sgoldstinos \cite{Gorbunov:2000th}.  With the current preliminary design \cite{Chou:2016lxi,Curtin:2017izq}, it is unclear whether MATHUSLA would have sufficient sensitivity to the $E_X\sim \order{100\mbox{ MeV}}$ particles coming from soft kaon decay products to uncover new physics there.  However, it is worth considering potential design modifications, such as additional tracking, calorimetry, or small magnetic fields, that would augment both low energy sensitivity and the ability to discriminate backgrounds in this region.  Given the exciting potential of MATHUSLA to access hidden particles and its ability to run symbiotically on LHC collisions evading the need for devoted beam time, further study of the costs and benefits of modifying the MATHUSLA design to maximize the sensitivity to low energy particles originating from rare meson decay is of paramount importance.  

\begin{acknowledgments}
We thank  J.~P.~Chou, D.~Curtin, S.~Gori, G.~Gollin, D.~Robinson, and J.~Shelton for useful discussions.  We especially thank J.~P.~Chou, D.~Curtin and J.~Shelton for comments on the manuscript.  We are especially grateful to S.~Knapen, M.~Papucci, and D.~Robinson for noting an error in the first draft, and for discussions about the SHiP acceptance.  This work was supported in part by DE-SC0015655.  This work was completed at the Aspen Center for Physics, which is supported by NSF grant PHY-1607611.
\end{acknowledgments}

\bibliography{MATHbib}

\end{document}